# Reverse Stress Testing for Supply Chain Resilience

Madison Smith[1], Michael Gaiewski[1], Sam Dulin[2], Laurel Williams[1], Jeffrey Keisler[3], Andrew Jin[4], Igor Linkov*[4]

[1]Credere Associates LLC, 776 Main St., Westbrook, ME 04092

[2]Northwestern University, 633 Clark St., Evanston, IL 60208

[3]University of Massachusetts Boston, 100 Morrissey Blvd., Boston, MA 02125

[4]Environmental Laboratory, US Army Engineer Research and Development Center, 696 Virginia Rd., Concord, MA 01742

## Abstract

Supply chains' increasing globalization and complexity have recently produced unpredictable disruptions, ripple effects, and cascading resulting failures. Proposed practices for managing these concerns include the advanced field of forward stress testing, where threats and predicted impacts to the supply chain are evaluated to harden the system against the most damaging scenarios. Such approaches are limited by the almost endless number of potential threat scenarios and cannot capture residual risk. In contrast to forward stress testing, this paper develops a reverse stress testing (RST) methodology that allows to predict which changes, with probabilistic certainty, across the supply chain network are most likely to cause a specified level of disruption at a specific entity in the network. The methodology was applied to the case of copper wire imports into the USA, a simple good which may have significant implications for national security. Results show that Canada, Chile, and Mexico are predicted to consistently be sources of disruptions at multiple loss levels. Other countries (e.g., Papua New Guinea) may contribute to small disruptions but be less important for the catastrophic losses of concern for decision makers. Other countries' disruptions would be catastrophic (e.g., Chile). The proposed methodology is the first case of reverse stress testing application in complex multilayered supply chains and can be used to address both risk and resilience.

## 1. Introduction

Modern economies depend on a vast web of production and distribution networks that move materials and goods around the world. Each link in these networks represents a dependency for inputs, logistics, and market access. The efficiency of global supply chains has enabled remarkable gains in productivity and specialization. Yet the same interdependence has also created structural fragilities that become visible only when disruptions occur. Events such as the COVID-19 pandemic, the semiconductor shortage, or the war in Ukraine have revealed how localized disturbances can cascade through multiple tiers of suppliers, amplifying their effects until entire sectors or regions experience economic stress [1,2]. Traditional methods for managing risk have struggled to keep pace with this complexity and the broad geographical distribution of supply chains.

The study of resilience offers one path to address this challenge. Resilience analysis examines how a system continues to function when parts of it fail, rather than preparing it to face specific threats. In this framing, the central question is not what disruption might occur, but how the structure and operation of the system influence its ability to absorb a disruption, then recover and adapt. This perspective is described as threat agnostic because it does not depend on an enumeration or characterization of specific hazards and vulnerabilities to assess risks [3]. For supply chains, adopting this perspective means looking beyond individual supplier failures or regional shocks to examine how losses propagate through the network [4]. Analytical tools must be able to identify vulnerabilities and response capacities without assuming which hazard will trigger them.

Stress testing has emerged as one of the most promising methods for examining the performance of complex systems under adverse conditions [5,6]. It allows analysts to explore "what if" questions and to compare system behavior under various hypothetical disruptions. The technique was first developed in the financial sector, where regulators sought to understand how institutions might respond to extreme market shifts or correlated losses. In that setting, a stress test applies a hypothetical shock to a modeled system and measures how its components respond. The method helps decision makers identify critical dependencies, assess the adequacy of buffers, and evaluate whether the system can survive under specified conditions [7]. Over time, this logic has been extended beyond finance to other sectors, including energy [8], logistics [9], and healthcare [10]. Each of these areas may be using stress testing to explore how operational or structural changes influence resilience (e.g. [11,12] and citations therein).

In supply chain management, stress testing initially took the form of structured workshops or scenario exercises in which managers discussed how disruptions might propagate through their networks [13]. These early applications were qualitative and depended heavily on expert judgment. They improved awareness of interdependencies but did not provide quantitative insight into how losses would evolve over time. The growing availability of trade and production data has since made it possible to formalize stress testing into computational models. Recent work has applied simulation and network analysis to test how changes in demand or supply at one node affect outcomes elsewhere in the system [14–16]. Such work has given rise to what can be called forward stress testing (FST), where analysts impose an assumed shock and observe its consequences.

Even though FST has attracted interest of the regulatory industry and academia as a promising approach for evaluating risk and resilience in supply chains [5], its strength is also its weakness. Every forward stress test depends on the choice of initiating threat events. In global supply chains, the number of possible disruptions is effectively limitless, ranging from natural disasters and labor strikes to geopolitical conflict and data failures. The analyst must decide which scenarios to model, leaving vast regions of potential risk unexplored. Even sophisticated probabilistic models cannot capture unknown or emergent interactions among suppliers, markets, and infrastructure. The result is a kind of tunnel vision: forward stress testing can illuminate only those vulnerabilities that fit within predefined scenarios. This limitation has led researchers to search for methods that identify fragility without requiring a prior specification of the shock itself [17,18].

Reverse stress testing provides a methodology to fill that gap (Figure 1). The concept originated in finance, where regulators used it to determine what kinds of conditions would push a bank or portfolio into distress. Instead of starting with a scenario and measuring its outcome, the analysis begins with an undesirable state (such as insolvency or a defined loss threshold) and works backward to determine what combination of factors could produce that result [19]. This inversion of logic enables reverse stress testing to discover rather than validate scenarios. It asks "what would need to happen for this outcome to occur?" rather than "what happens if this event takes place?" The approach therefore aligns naturally with resilience analysis, which focuses on the internal dynamics of systems and their ability to maintain function.

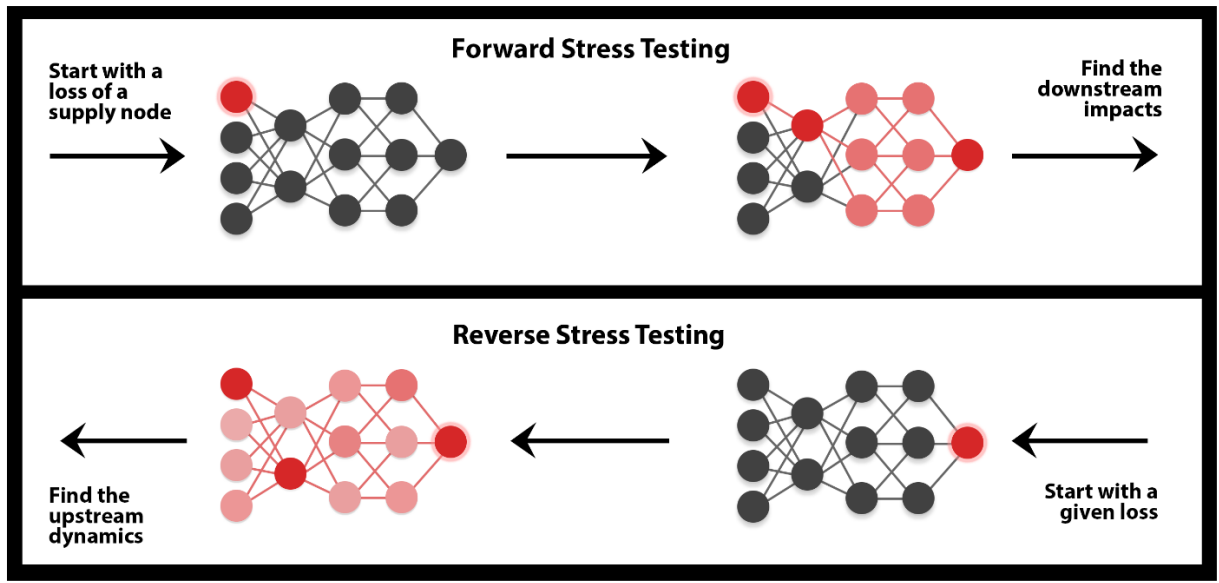

Figure 1: Comparison between general workflows for the forward and reverse stress testing of supply chains.

Despite its conceptual appeal, reverse stress testing has remained largely confined to financial applications [7,20,21]. The main barrier has been the difficulty of translating it from relatively easy modeled financial assets into multidimensional and complex physical or operational systems. Working backward from an outcome involves solving an inverse problem in which multiple combinations of disturbances can produce the same observed effect. For complex systems, the number of possible configurations is enormous. In supply chains, for example, a reduction in final product availability could result from many different combinations of supplier delays, transportation bottlenecks, or market contractions. Identifying the most probable combinations requires not only detailed data but also probabilistic methods capable of handling uncertainty and interdependence.

This paper proposes an RST methodology able to draw the most probable scenarios from a seemingly infinite scenario space to deliver reliable data-driven insights into the fragilities of the complex system to the end user. The methodology requires a reinterpretation of some of core concepts derived from financial industry. In finance, the entities of interest are banks and portfolios, and the metric of concern is capital adequacy. In supply chains, the entities are suppliers, manufacturers, and trading partners, and the relevant metric is the continuity of goods or materials through the network. An RST begins by defining an outcome of concern, such as a specified reduction in supply to an end user. The RST then estimates which upstream changes in production, capacity, or trade could plausibly produce that outcome. Because these systems are layered, the analysis must account for how losses at one stage translate into constraints at others. The task is therefore to identify the configuration of upstream perturbations most likely to generate the observed downstream disruption.

Ultimately, this approach reframes the analysis of supply chain resilience. Instead of beginning with hypothetical hazards, it begins with outcomes that decision makers care about and then uncovers the internal dynamics that could lead there. In doing so, it shifts attention from prediction to inference, from defensive preparation to structural understanding. As economies continue to rely on globally distributed production systems, the ability to anticipate where and how disruptions can accumulate will determine not only operational continuity but also the stability of national and international markets. Reverse stress testing provides a systematic way to obtain that insight.

This study demonstrates application of an RST methodology using a real-world case – the global copper supply chain and its impacts on copper wire imports into the United States (US). United States relies on the import of copper goods, such as copper wire, to maintain national security priorities (The White House, 2025). Copper has been designated a critical mineral by the US Department of Energy (DOE) since 2023, as it plays a major role in the development of energy systems [23,24]. Therefore, maintaining a constant stream of copper goods into the US can be seen as an important goal to maintain critical infrastructure systems. For example, Chile is home to major copper mines and processing facilities that produce and export the majority of copper ore, concentrate, and refined copper goods to other parts of the world [25]. Other important countries in the production of copper include Zambia (unrefined copper) [26], Democratic Republic of the Congo (both unrefined and refined copper) [27], and Japan (refined copper) [28]. The United States is a major importer of copper wire (over $2B in import in 2023 alone) from countries like Germany [29], United Arab Emirates [30], and Canada [31]. This global supply chain encompasses many redundancies and loops that prove to be a complex supply network even for simple goods such as copper wire. Due to this complexity, fragilities and bottlenecks are not expected to be obvious in all cases. Therefore, this paper applies reverse stress testing to understand where problems may occur in this global supply chain. Reverse stress testing is used to identify the most likely scenarios that could lead to small (5%), medium (20%), and large (50%) drops in observed copper wire imports to the United States. Critical countries in the trade network of goods earlier in the supply chain are identified that may signal potential future disruptions in the US's copper wire imports.

# 2. Results

## 2.1. Supply Chain Network for Copper

The supply chain network for copper is constructed from the Comtrade data as described in Section 4.2 [32]. As the network used in this use case was constructed to focus on import of copper wire only into the United States, the quantities of goods flowing through each layer of the supply chain is expected to taper off as the transactional relationship between entities travels from raw materials to the final good (Figure 2, left panel).

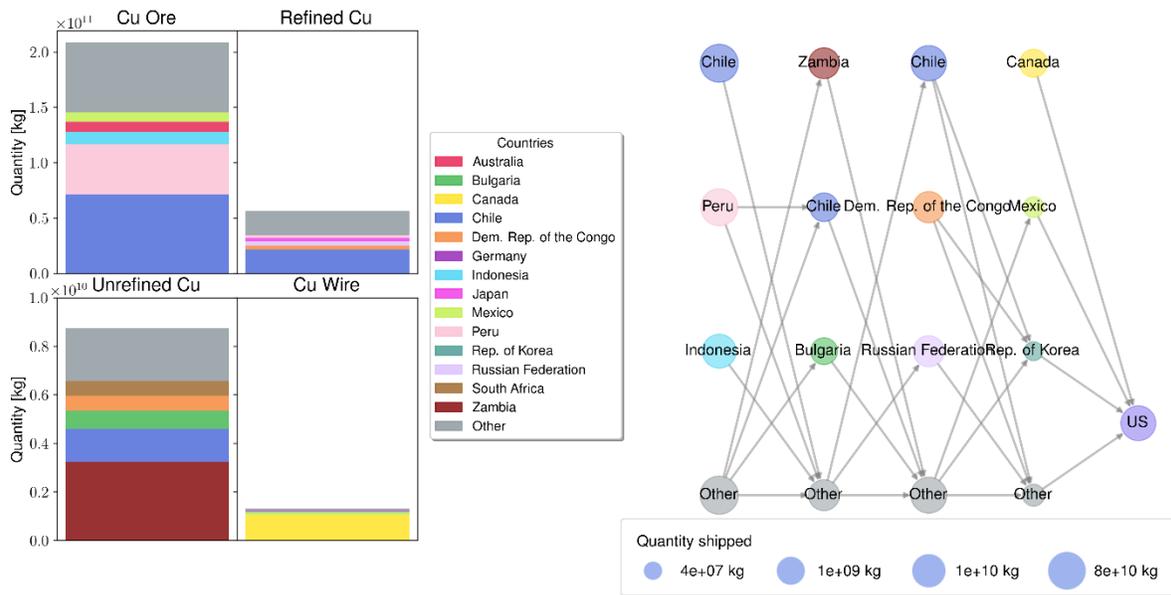

*Figure 2: (Left) Quantities flowing through each layer of the constructed network. (Right) Constructed supply chain topology based on the Comtrade data [32]. Countries outside of the top five and three contributors are aggregated into other for the left and right figures, respectively.*

Another key characteristic of the constructed network is its topology (Figure 2, right panel). For each layer in the network, the three major contributors to the production of each good is shown in addition to its relationships to the next layer. For most of the entities represented in the constructed network, multiple relationships with other countries are captured, both as incoming and outgoing shipments of goods. The exceptions to this trend are for the Democratic Republic of the Congo's production of unrefined copper and for Canada's production of copper wire. These two nodes in the constructed network have no observable imports of other goods from preceding layers. This topology may be due to two reasons; first, the data records for imports may simply be incomplete for these exact goods and countries. Second, the production of intermediate goods may happen within the country itself, making the import/export trace of these goods nonexistent in the current methodology. For example, if Canada has the capability to refine copper within its borders, there will exist edges tracing imports of copper ore and exports of copper wire. These dynamics drove the development of the reserves mechanic discussed in Section 4.1.1 and the Appendix.

## 2.2. Most Probable Supply Chain Disruption Scenarios

Table 1 contains results for the most likely scenarios computed by the reverse stress testing method. Results are limited to the top five most impactful countries at each layer for each shock scenario. The results suggest certain countries are key players in the supply chain, with predictions consistently identifying them as sources of disruptions at multiple loss levels. For copper ore, Turkey, Spain, and Brazil are consistently the top three most likely causes for concern. For unrefined copper, Chile, Bulgaria, Sweden, and Germany appear most often as the likely causes for concern. For refined copper, the United States, Germany, and Chile appear most frequently. On the other hand, other countries only appear in more specific scenarios. For example, Zambia does not appear until the 20% loss level for unrefined copper and Belgium only appears in refined copper at the 5% loss level.

The US itself appears as a cause for concern at intermediate stages of copper goods. The copper wire market in the US depends to a degree on its own ability to move copper at intermediate stages to other countries. While this dynamic is clear from the topology of the modeled supply chain, it is also notable the degree to which it is predicted to be important at these intermediate layers, suggesting an avenue for internal management of potential disruptions.

Table 1: Deterministic prediction of top five largest losses across the supply chain most likely to cause a 5%, 20%, or 50% expected loss in copper wire goods entering the US.

|  | Copper Ore | | Unrefined Copper | | Refined Copper | | Copper Wire | |
| --- | --- | --- | --- | --- | --- | --- | --- | --- |
|  | Country | Qty (kt) | Country | Qty (kt) | Country | Qty (kt) | Country | Qty (kt) |
| 5% Loss | TUR | 0.04 | CHL | 0.16 | USA | 0.72 | CAN | 4.39 |
|  | ESP | 0.04 | BGR | 0.12 | DEU | 0.31 | MEX | 0.73 |
|  | BRA | 0.04 | DEU | 0.08 | BEL | 0.20 | KOR | 0.64 |
|  | USA | 0.03 | SWE | 0.08 | JPN | 0.20 | CHL | 0.58 |
|  | GEO | 0.03 | USA | 0.05 | CHL | 0.17 | PER | 0.56 |
| 20% Loss | BRA | 0.12 | CHL | 0.47 | USA | 3.08 | CAN | 26.45 |
|  | TUR | 0.10 | BGR | 0.31 | DEU | 0.78 | MEX | 3.45 |
|  | ESP | 0.09 | SWE | 0.20 | CHL | 0.71 | KOR | 2.88 |
|  | USA | 0.07 | DEU | 0.19 | COD | 0.46 | CHL | 2.47 |
|  | GEO | 0.07 | ZMB | 0.12 | JPN | 0.41 | PER | 2.33 |
| 50% Loss | BRA | 0.18 | CHL | 0.78 | USA | 7.68 | CAN | 84.57 |
|  | TUR | 0.14 | BGR | 0.46 | CHL | 1.76 | MEX | 8.87 |
|  | ESP | 0.13 | SWE | 0.28 | COD | 1.39 | KOR | 6.99 |
|  | GEO | 0.11 | ZMB | 0.27 | DEU | 1.09 | CHL | 5.65 |
|  | MKD | 0.11 | DEU | 0.23 | JPN | 0.75 | PER | 5.20 |

*Quantities are measured in kts and country abbreviations follow the ISO-3166-1 3-letter country codes. Cells are colored based on the magnitude of the predicted production loss.*

The most important anticipated changes in quantities of goods moving through the system for the 5%, 20%, and 50% anticipated shocks share notable similarities. Canada, Mexico, Chile, Republic of Korea, and various European countries, for example, are key players in the sustainable pipeline of copper goods into the US as seen here.

## 2.3. Predicted Losses Across the Supply Chain

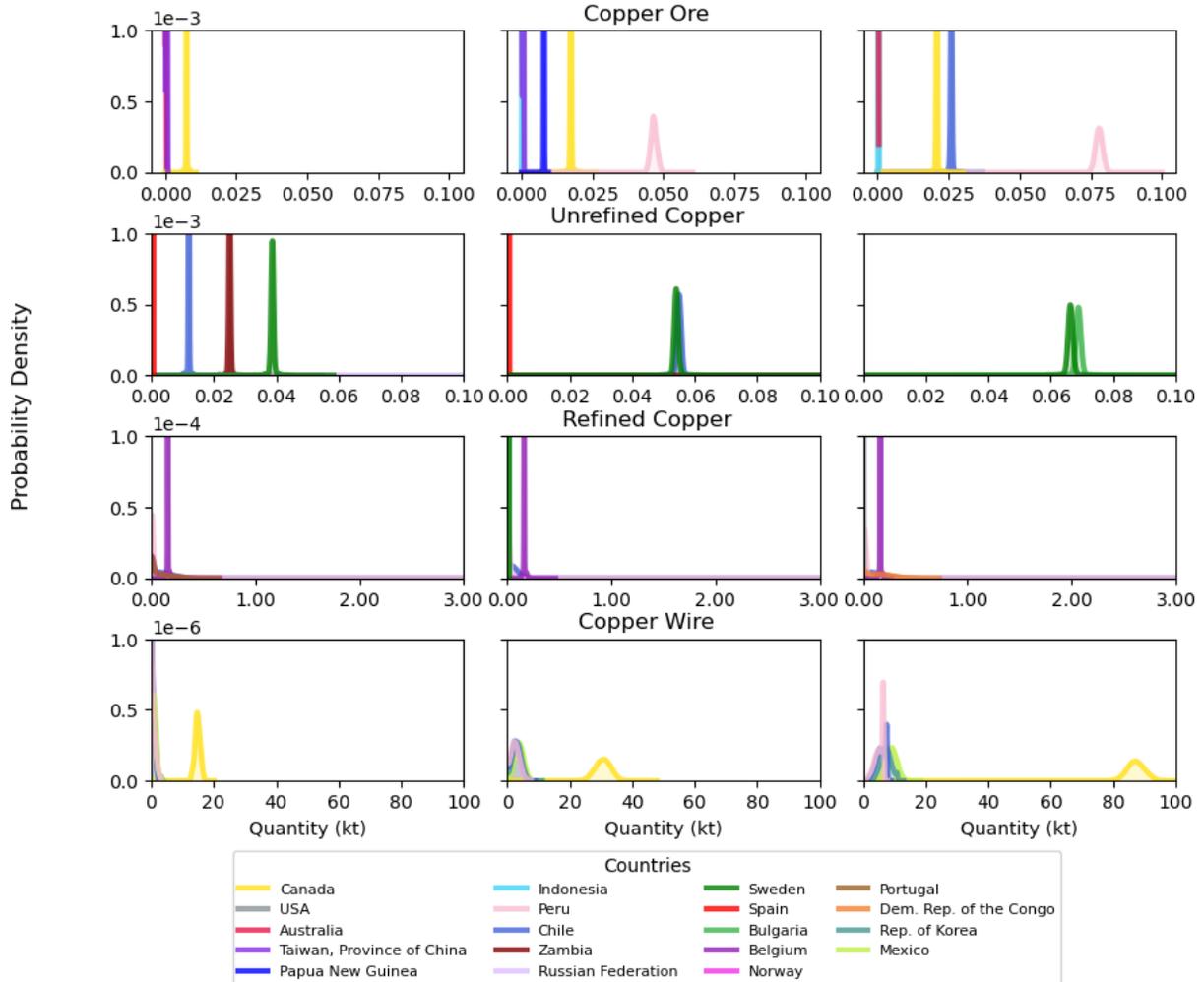

*Figure 3: Probability density functions for predicted losses across the supply chain, disaggregated by sourcing country. PDFs are shown for the five main predicted contributors to a loss at each layer. Each column represents a different scenario (5%, 20%, and 50% reductions in US imports of copper wire).*

Figure 3 shows, for top five predicted contributors, the constructed probability distributions functions (PDFs) of possible production loss at each supply chain tier that is predicted to result in the US copper wire import loss scenarios of 5%, 20%, and 50%. For example, within the copper ore tier, Canada, the USA, Australia, Taiwan, and Papua New Guinea are predicted to be the most likely to contribute to a resulting 5% loss of copper wire to the end consumer (the United States), with relative probabilities represented by peak heights of approximately 0.001, and quantities of between 0 and 0.01 kT. The back propagation technique asks what loss at the preceding layer could cause the specified configuration of loss at the layer in question: therefore, the results for each tier and a given loss scenario can be interpreted independently – what configuration of loss in this specific tier could cause a loss of % of copper wire to the United States? – or as a general description: how could losses potentially propagate through the supply chain's tiers to result in this loss scenario?

As with the deterministic results described above, some countries are predicted to play significant roles for multiple scenarios and/or tiers, e.g. Chile and Peru. The top five countries as displayed in the PDFs per scenario and tier vary from those displayed in Table 1. These variations reflect higher stability resulting from the sampling method used to create the distributions versus the Table 1 deterministic results, which are by nature more sensitive to the covariance matrix.

As the shock scenarios in the experiment increase in magnitude, the associated predicted losses for each supply chain tier increase correspondingly. Predicted losses for the ore tier vary between 0 and 0.3 kT, the predicted losses for the refined copper tier reach 0.5 – 1.0 kT, and the predicted copper wire loss quantities are one to two orders of magnitude higher again.

Some countries emerge as notable players for predicted quantities lost within a given tier as the loss scenario magnitude increases. For example, at the copper ore tier, the Peru and Chile peaks move to the right with scenario severity, indicating increased quantity of ore predicted to be contributed by both to the resulting loss. Similar behavior can be seen for the Sweden peak in the unrefined copper tier and the Canada peak in the copper wire tier.

In other cases, "major player" behavior for a given tier remains fairly consistent throughout scenario severity: for example, Belgium is present at about the same quantity for all loss scenarios in the refined copper tier. Finally, as with the deterministic results described above, some countries only appear as a top five contributor for more specific scenario types. For example, Papua New Guinea is only predicted to play a role in the less severe raw ore tier scenarios.

The height of the peaks in Figure 3 represents higher probability of occurrence – in other words, that a predicted loss by a predicted country is more likely to be involved in the loss scenario. Peak height also demonstrates relative differences between tiers of the supply chain and shock scenarios. Generally, probabilities are highest for predictions at the higher tier, reaching a maximum of 0.001, and decrease by orders of magnitude progressing through the tiers, indicating decreasing certainty as the tiers progress toward the end user.

When interpreted in conjunction with the width of a peak, the height represents the relative likelihood that a country will play a role in the specified loss over other countries in that layer. Narrower peaks suggest that past behavior of these entities in the network is less uncertain. Peaks are both generally narrower and highest at the copper ore tier and generally become broader and smaller progressing toward the copper wire tier. Narrower and higher peaks in these PDFs represent more certain and more probable losses, respectively.

Similarly, progressing toward higher end loss scenarios, the full width half maximums of the PDF peaks increase. In other words, as losses at the end node in the network increase (i.e., experimental shocks going from small to large losses), it is more likely to predict that the observed loss could have come from several different countries across the supply chain and not just one.

# 3. Discussion

Currently supply chain risks are assessed through supply chain illumination and assigning metrics associated with categories of risk (i.e., geopolitics in the country of origin, business records, exposure to natural disasters, etc.). If more advanced modeling is done, forward stress testing is applied where major catastrophic scenarios are considered, and components failure is recorded and managed. Application of reverse stress testing shifts analyses from studying hypothetical disasters toward the examination of conditions that make loss possible. It asks what set of disruption and resulting changes in a supply chain, perhaps small on their own, could combine to create cascading failures. By working backward from specified failures thresholds, the method exposes the hidden structure of system vulnerability that connects suppliers, producers, and consumers across interconnected supply chain and logistics networks.

The results for the copper wire supply chain suggest that there are certain countries that indeed are key players and are predicted to consistently be sources of disruptions at multiple loss levels. The results are consistent with what is found in the literature -- Canada, Mexico, Chile, as well as several European countries, seem to play the biggest roles when it comes to potential US copper wire losses (e.g., [33]). These countries emerge in the results as key players because they produce a majority of these goods in our modeled supply chain. For example, the prediction of production losses across the copper wire tier of the supply chain (Figure 3) shows large values of predicted losses for Canada. This result can be tied back to the US's exposure to Canada in its copper wire supply chain (Figure 2); because most of the US's import of copper wire comes from Canada, disruptions in Canada's production of wire directly impacts the US.

While the importance of key players in copper production for the US copper wire import can be derived based on literature and common sense, it is also important to note the less obvious dynamics unveiled by application of RST. Countries that are likely to contribute to overall losses during small disruptions may be less important for a high disruption threshold (e.g., Papua New Guinea is important for the threshold of 5% and 20% but less important for the 50% threshold). Situations where a country only becomes a possible major contributor at higher disruption scenarios are also observed. Peru becomes a major contributor at the major (50%) disruption threshold but is not in the top five contributors for the smaller thresholds This is likely due to it being a major source of copper ore, and a major disruption at that layer can cascade through the supply chain and result in a major disruption at the finished good tier.

In addition to the changes in results across disruption thresholds, the RST results also shift importance away from stable key players to the less stable, smaller configurations that together become a larger problem. For example, the PDFs for copper ore production (bottom row Figure 3) across all three scenario experiments do not show Chile as the highest contributor to the predicted production loss, even though it is the major contributor to copper ore in the modeled supply chain (Figure 2). These results challenge how managers and policymakers have been taught to think about supply risk. The instinct is to focus only on the big players: the dominant producer, the key logistics hub, the one supplier everyone depends on. Protecting those nodes seems like common sense. However, the reverse stress testing results show that systemic losses can just as easily come from the alignment of many small and medium contributors

The probabilistic framing of considering potential losses across a supply chain is an important choice for direct application in supply chain management spaces. This is highlighted in the difference of results between Table 1 and Figure 3. Relying on the most probable point estimates generated by the method does not take into consideration the many sources of uncertainty—both aleatoric and epistemic – in this use case. The covariance matrix is derived from a limited set of observations, driving a key source of the uncertainty we can see in the results. When the covariance matrix is perturbed and sampled, a more complete picture can be drawn about the underlying distribution of production capabilities for each layer in the supply chain.

Policymakers looking to use data-driven methods for their decision making can use the RST method to gain insight across the many tiers and dynamics present in a supply chain. For the more obvious dynamics, such as those seen for Canada in the copper wire tier in this case study, RST provides reinforcement to the importance of exposure to risk in a supply chain. On the other hand, the less obvious results may challenge how policymakers and managers have been taught to think about supply risk. The instinct is to focus only on the big players: the dominant producer, the key logistics hub, the one supplier that everyone depends on. Protecting these nodes seems like common sense. However, the RST results show that systemic losses can just as easily come from the alignment of many small and medium contributors. For examples, the PDFs for the predicted production loss of copper ore (Figure 3) across all three disruption scenarios do not show Chile as the highest contributor to the predicted losses, even though it is a major contributor to copper ore in the supply chain (Figure 2). Instead, the results show several other countries contributing small amounts to the loss that sum to an overall considerable loss in production. The RST method provides insight into these more nuanced dynamics and presents them in a probabilistic manner for policymakers and managers to assess directly.

Even though both FST and RST frameworks provide insights into how a supply chain may fall apart, reverse stress testing answers a different set of questions about the conditions under which a supply chain may degrade. While FST can build a catalog of "what if" scenarios for managers based on designed shock scenario simulations, RST averts the burden of scenario design and provides a threat-agnostic way to view a supply chain's resilience. RST focuses on the end user's concerns without being limited by the imagination of what could go wrong. It directly searches the infinite space of "what ifs" and finds the configurations and dynamics that can lead to untenable disruptions. Because of its ability to identify probable sources of disruption before they happen, RST more directly assesses the risk and resilience of a supply chain. Together, the forward and reverse forms of stress testing provide a more complete toolkit for managers to prepare for, absorb, recover from, and adapt to supply chain disruptions.

As this is the first application of reverse stress testing to supply chain management, some simplifications and assumptions were made leading to several limitations to the model. The model relies on the assumption that the values being analyzed are multivariate normal distributions centered at zero. This assumption holds loosely in the current application of the equations, but deriving alternative distributions may be possible and necessary. Another assumption made is that all relevant information to describe the way supply chains operate is encoded in the historical transactional data. Additional factors, such as geopolitical scenarios, may have direct impacts on the estimated covariance matrices and flow of goods. These limitations serve as avenues for future research and addressing them would ultimately help create a more comprehensive RST model.

This paper and its framework for RST provide a baseline for future quantitative and threat-agnostic methods to continue to explore these concepts in complex systems. Other modeling approaches such as network science and agent-based modeling can be layered into this framework to provide the extra details needed for complex decision making. In its current state, the method operates as a companion tool for users with supply chain subject matter expertise to make better-informed, data-driven decisions. Anticipating where compounding and cascading events can come together to create adverse impacts, RST provides a probabilistic framework to answer more than "where is the risk". Supply chains do not operate on certainty; they operate on distributions of outcomes. Understanding those distributions helps managers see which configurations of suppliers or trade routes bring the system close to a tipping point. It also shows where small adjustments (e.g., shifting one input, changing timing, adding slack) can move the entire system back into a safer zone. As more real-world complexities such as circular processes, socio-behavioral dynamics, and policy constraints are incorporated into this type of tool, decision makers will have deeper insight into what makes a supply chain resilient.

# 4. Methods

The proposed methodology constructs a network of entities linked by trade. Each layer represents a stage in production, from raw material extraction to final manufacturing. Statistical models estimate how quantities change across layers based on historical data, and Bayesian sampling techniques capture uncertainty in those relationships. The method then identifies the combinations of upstream losses that correspond to a target reduction in output at the end of the chain. Specifically, this method seeks to conceptualize a supply chain so that undesired losses can be propagated through the network and predict the most likely scenarios that could have caused the undesired loss, as well as the entities that are most likely to contribute, throughout multiple likely scenarios, to a specific loss in available resources for an end user.

## 4.1. Generalized Framework for Reverse Stress Testing

The generalized reverse stress testing follows a five-step process to generate the results (Figure 4). Step 1 constructs the network as an abstraction of the true complex relationships seen in the supply chain. This involves determining the input materials needed for a final good, and sorting each entity into a corresponding layer based on its production of the identified goods. It further requires the network topology of transactional relationships to be defined via observational evidence. Step 2 is a single layer reverse stress test, wherein a loss threshold is defined at the end user and the sources of that loss are predicted at the previous layer based on the end user's direct suppliers. This step is akin to the reverse stress testing on financial portfolios. Step 3 backpropagates these loss predictions through the entire layered network. This step is the important extension that allows for previous reverse stress testing methodologies to be applied to supply chains. Step 4 repeats Step 2, but produces other less likely, but still probable, loss scenarios for each layer in the network. While the most likely scenarios produced by the first run of Steps 2 and 3 are important, this methodology is intended to account for other scenarios that may be possible. Step 4 accounts directly for those other scenarios that may play an important role in the supply chain's overall resilience. Finally, Step 5 aggregates the predicted loss scenarios in a probabilistic manner for output. This step uses Bayesian methods to collect large samples from the true underlying distribution of probable losses and accounts for the uncertainty present in the observed relationships between entities.

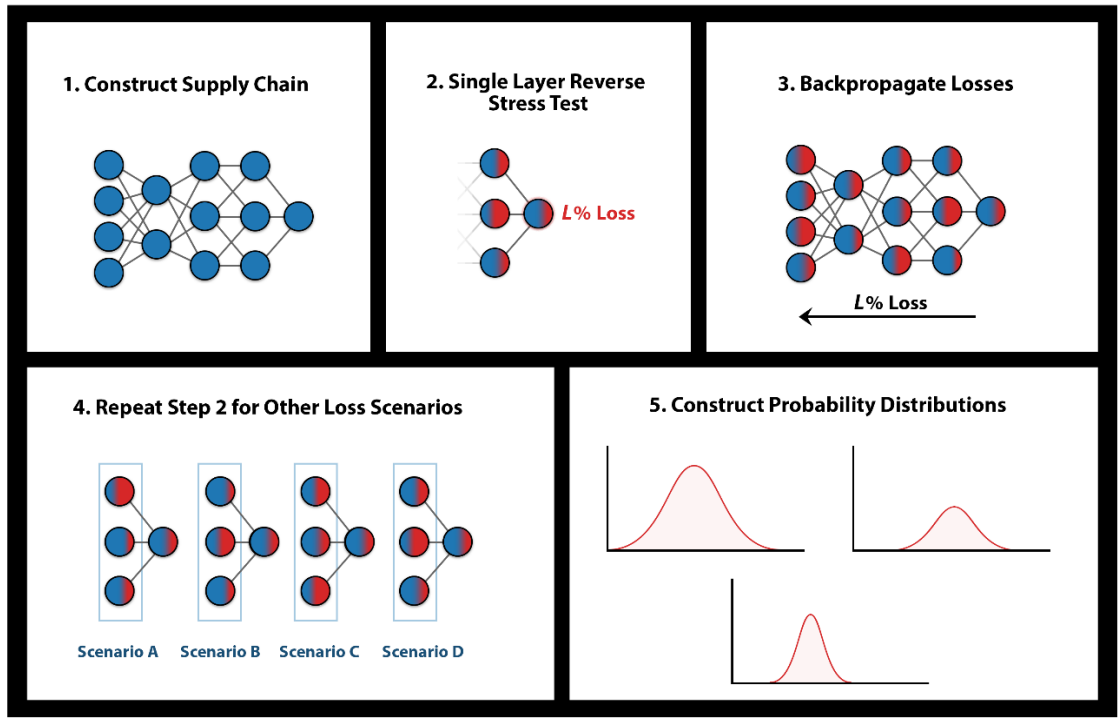

*Figure 4: Methodology workflow from network construction through to the predicted losses for each entity in the network.*

### 4.1.1. Network Construction

The supply chain is assumed to be multilayered where each layer represents a transformation from one good type to another. Goods are assumed to be ordinal and progress between layers until the end good (i.e. $T_1 \rightarrow T_2 \rightarrow \ldots \rightarrow T_M$). Goods cannot be reversed (e.g. a metal cannot be recycled or turned back into ore). Only historical transactions between entities are considered, and entities are assumed to import one material and export the next ordinal material (i.e. an unrefined metal to a refined metal). Each node in the network represents an entity. Edges in the network represent directional transactions from one entity to another.

The construction of the network starts at the end layer $T_M$, which contains only one entity. Only entities that directly export to layer $T_M$ are considered final product to the end node are added to layer $T_{M-1}$. Similarly, only entities that export goods to some entity in layer $T_{M-1}$ are added to layer $T_{M-2}$, and so on.

In existing supply chain data, flows may not always align with expected movements through the network. As goods move from raw materials to the finished state, overall weight is expected to change. To ensure functionality of this model, a "reserve" system is implemented. When there is a mismatch between input and output weights, the model incorporates how much an entity may be holding in reserve. This technique helps with the potential incompleteness of the given raw data. A full description of this reserve approach can be found in the Appendix. In a closed and complete system, the back propagation model would show the same total quantity losses at each layer.

### 4.1.2. Single Layer Reverse Stress Test

For a target node $n_j$, a covariance matrix $D_j$ is constructed from historical monthly percentage changes in exports between its $N$ immediate suppliers. Because some suppliers are highly transient, several assumptions and transformations were developed to make a more accurate covariance matrix for this use case. (See Appendix). The distribution of changes in quantity is assumed to be multivariate normal with mean zero.

The RST of a supply chain is made possible through adaptation of methodologies used in the financial sector [21]. Generally, under this methodology, the mean scenario (i.e. most likely) is the desired loss, $L$, multiplied by row-wise sum of the scaled covariance matrix (Equation 1).

$$a_j = \frac{L}{\sum_R \sum_C D_j} D_j \vec{1} \tag{1}$$

Each other scenario is derived by shifting the mean scenario a set distance (see the parameter $q$ in [21]) away along the principal components of the covariance matrix transformed onto the hyperplane of all scenarios that equal an $L$ loss (or gain). The relative likelihood of that scenario is a function of the eigenvalue associated principal component. Further details on the method, as well as any modifications made to the method are discussed in the Appendix. A feature was added to ensure that no predicted transaction can be below zero.

### 4.1.3. Back Propagation Model

Nearly all supply chains have multiple "layers" or "tiers" comprised of entities that execute the various sequential intermediate steps of transforming a raw material into a consumer good. So far, the methods described have been establishing the prediction of changes in $\Delta Q$ in a single layer of the supply chain. However, these concepts can be expanded to consider how an $L$ gain or loss at a consumer node could come from many upstream suppliers in a multi-tiered supply chain. If the order of good types from raw materials to final products is known, one may be interested in predicting changes at each layer from the consumer node to the raw materials from running the reverse stress test.

To implement this prediction, the reverse stress test is run for each consumer node and the most likely predicted changes in quantity are applied to that entity's immediate suppliers. At each immediate supplier a $K$ kg gain or loss in output could potentially be explained by a $K$ kg loss in input from *their* immediate suppliers of the previous intermediate good. One would calculate what $L$ corresponds to a $K$ kg loss and the reverse stress test can be run again. This process is repeated until all changes are tracked from the consumer node to the applicable entities at the beginning layer of the network, or the entities that supply the raw materials. This iterative model is referred to as the *back propagation model*, tracing the most likely cause of losses throughout the tiered supply chain. One feature is that multiple changes can be made to one upstream entity if it supplies multiple immediate downstream entities. The changes are summed together before computing the new $L$ and running the RST on this entity's immediate suppliers.

### 4.1.4. Probability Distributions for Likely Production Losses

Probability distributions to anticipate the possible range in expected output quantities ($Q_i^j$) were constructed using a Bayesian approach. Since the covariance matrix is always symmetric positive definite, one can draw $M$ random samples of a perturbed covariance matrix using an inverse-Wishart distribution [34]. The reverse stress test can be run again on each of the $M$ perturbed covariance matrices. As a result, there are $M*(N+1)$ weighted scenarios. These scenarios are combined into a weighted ensemble for each country and good pair, where the weights are determined by the relative likelihood of each scenario compared to its most likely scenario produced by each covariance sample. For the plots seen in Section 2, $M = 1000$ and the resulting samples were smoothed via kernel density estimation to create the probability density functions. In some cases, entities will have only one supplier, so any losses are guaranteed so come from that single supplier, and therefore, the results of the RST are deterministic. To provide a probability distribution as output in this case, a Dirac delta function at the predicted quantity loss is used in lieu of a smoothed kernel density estimation from sampled scenarios.

### 4.2. Use Case: Copper Wire in the US

The methodology is illustrated for the case of copper wire in the US. The process of making copper wire is simplified to four main stages: raw copper ore, unrefined (blister) copper, refined copper, and, finally, copper wire. This is a simplification of the actual copper wire supply chain, eliminating known byproducts, added materials, and recycling [35]. These simplifications allow for the capture of major stages for copper wire production without adding complexity to the network topology and interpretation of RST results.

Data was sourced from the United Nations Commodity Trade Statistics Database (UN Comtrade) for 4 HS Codes: HS2603 (copper ore/concentrate), HS7402 (unrefined or "blister" copper), HS7403 (refined copper and copper alloys), and HS7408 (copper wire) [32,36,37]. Import and export records between January 2018 to December 2024 were pulled from the database for all countries that contained one of the four HS codes identified above. The quantity of trade value was measured in kilograms per month calculated from Cost, Insurance, and Freight (CIF) records based on the price of the transaction and the spot price of copper. For the purposes of analysis, each country-good pair is a unique entity. Further information on how the data was cleaned is in the Appendix.

From this data, the supply chain network model was developed and applied for different disruption scenarios with different tolerance thresholds over the same timespan (12 months). The first scenario is a slight perturbation in the yearly received copper wire by the US, with an expected loss of 5%. The second scenario is a more moderate change received in copper wire goods at a loss of 20%. The third scenario is a more catastrophic change in received copper wire goods at a loss of 50%. All scenarios used the same $q$ value of 0.5 [21] when finding the "less likely" predictions.

# Data Availability

Data was sourced from UN Comtrade, however the cleaned version of the data is available upon request [32].

# Acknowledgements

The authors would like to thank the DARPA Resilient Supply-and-Demand Networks (RSDN) program for funding this research.

# Author Contributions

M.S. contributed to the methodology, data curation, writing – original draft, writing – review and editing, and visualization. M.G. contributed to the methodology, writing – original draft, writing – review and editing. S.D. contributed to the conceptualization, methodology, writing – original draft. L.W. contributed to the conceptualization, writing- original draft, writing – review and editing. J.K. contributed to the conceptualization. A.J. contributed to the conceptualization, visualization, writing – original draft, writing – review and editing, and supervision. I.L. contributed to the conceptualization, supervision, and funding acquisition.

# Competing Interests

# Materials & Correspondence

Corresponding author is Igor Linkov (Igor.Linkov@usace.army.mil)

# Appendix

## Data Cleaning

Extra data cleaning steps were needed before the UN Comtrade data was able to be transformed into a supply network of shipping goods between countries. Several steps recommended in [38] were used to limit the effects of outliers, missing data, and mismatched bilateral trade records. First, unit price was calculated from the value and quantity variables captured for each transaction record. A conservative upper percentile cut of 99% was used to identify and to remove only the most extreme outliers. Missing values were then imputed using the global median per each good type. Because of the sampling method, this dataset only captures import records, and therefore, bilateral asymmetries are ignored. More advanced techniques for handling the outliers [39], missing values [40], and bilateral asymmetries [41] can be used when the fidelity of the transaction data is important. As this data is used in a proof-of-concept use case, these simple data adjustments are sufficient. To further down select for the purposes of this paper trade relationships in the dataset were limited to those with more than 72 transactions over the 6-year period, or more than average of one transaction recorded per month. After these adjustments were made to the data, there was a total of 141,204 transactions between 125 countries.

## Reserves

In most transaction data sets, something akin to the following scenario may occur:

> Site A give 500kg of Good 1 to Site B at timestep 1. Site C give 200kg of Good 1 to Site B at timestep 1. Site B gives 1000kg of Good 2 to site D at timestep 2.

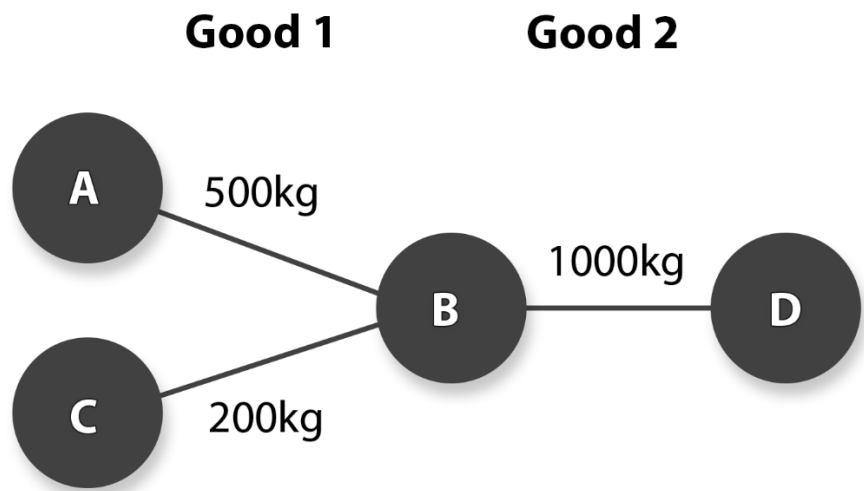

Figure 5: Example where flows do not align with expected good flows.

Based on the transaction data, Site B is only expected at this time to have 700kg of Good 1 to transform into Good 2 and ship to Site D; yet it shipped 1000kg of Good 2 to Site D. To address this situation, the concept of *reserves* (warehoused or stockpiled materials) is introduced. It is assumed Site B must have taken 300kg from its reserves to complete this transaction. A list of reserve transactions are created and added to the transaction list to make all transactions feasible. In other words, the reserve, if necessary, of a site is considered as an immediate supplier for running the reverse stress test. The example above would be modified to:

Site A give 500kg of Good 1 to Site B at timestep 1. Site C give 200kg of Good 1 to Site B at timestep 1. Site B transforms 300 kg of Good 1 into Good 2 from its reserve of Good 1. Site B gives 1000kg of Good 2 to site D at timestep 2.

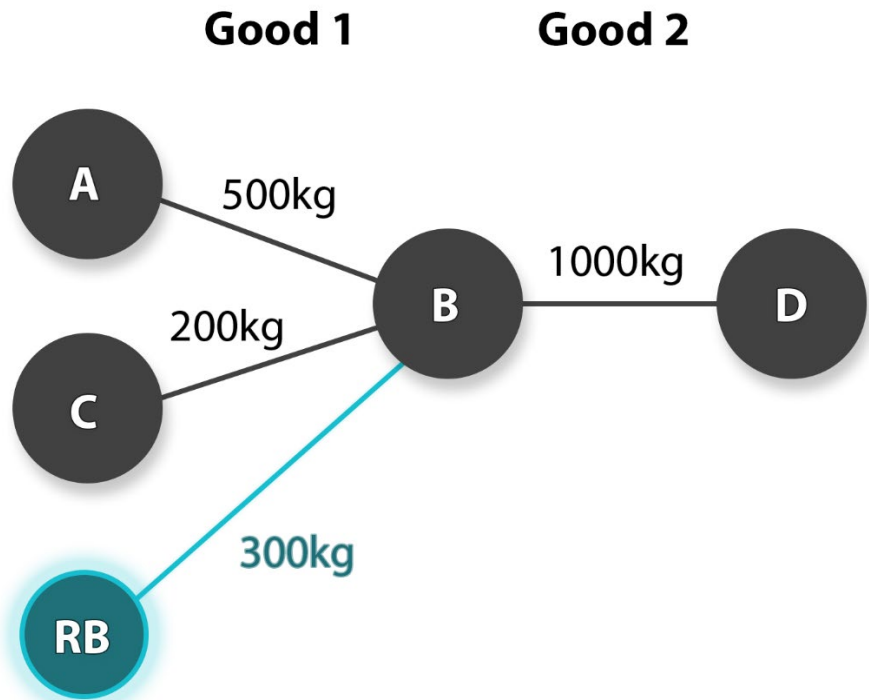

*Figure 6: Example that corrects good flows using reserves.*

In implementing this approach, the process of computing the various $L$ gain or loss values at each layer is now guaranteed to flow as expected. One limitation, however, is that by using reserves, it does not guarantee that there will be a consistent $K$ kg loss at each layer. This is because it is assumed that "reserve entities" have no immediate suppliers and they may occur at middle layers of the supply chain. Another limitation of this model is that if a transaction occurred that did not need to use any reserves, the corresponding reverse stress test will not consider using reserves as immediate suppliers. This phenomenon is seen because if a supplier gave nothing to that recipient initially, then any percentage change predicted by the reverse stress test of zero is still zero.

This reserve approach cannot fully address data discrepancies of this type. For example, the same phenomenon could result from missing entities that have transactions with entities represented in the data set. Such discrepancies can arise from inherent data structure (e.g., data sets that omit certain countries), variability in record-keeping between different contributors to the data set, methodological choices (e.g. exclusion of recycled and substitute inputs), and so on. The reserve model is realistic, as almost any entity can be assumed to keep at least some reserves and can imperfectly compensate for other data discrepancies because it builds in missing quantities, even if it does not specify where they specifically come from. For the purposes of the use case described, the reserve model reasonably approximates missing flows. Future improvements on the model could incorporate additional methodologies for data cleaning, validation, and modeling.

## Covariance Matrix

The calculation of a realistic covariance matrix, $D$, is crucial for the adaptation of [21]. For each transaction in the dataset, there is a timestamp associated with the time that the import was recorded. This time is assumed to be approximately associated with the time the shipment of goods arrives at the receiving entity, aggregated to a monthly resolution.

To construct the covariance matrix, it is ensured each pair of entities that provide goods to the same target entity have the same number of transactions. The user inputs a parameter $B$ and the algorithm sorts and sums the transactions into $B$ evenly spaced time intervals relative to the maximum and minimum time stamp of the given transactions. Sometimes, a time interval will have a sum of zero, which means that no transactions were completed during this particular time window. More research can and should be done to determine these effects, as well as the effects of different sized $B$ values. For the results in this paper, $B$ was set to 5.

The result is an $N$ x $B$ array of data, where $N$ is the number of unique suppliers. The two sets of transactions are multiplied by $B$ to help preserve the relative covariances, though this is an optional operation. The percentage changes between each of the first $B - 1$ columns with its adjacent column on the right are calculated. The resulting array $C$ is now an $N$ x $(B-1)$ array. To ensure that the covariance matrix does not have too large a magnitude, the percentage changes are clipped to be between -100 and 100. To finally calculate the $N$ x $N$ covariance matrix, the sample covariance formula is used as follows:

$$D = \frac{CC'}{B-1}$$

Additionally, while these covariance matrices are, by definition, symmetric positive definite, additional steps are sometimes needed to ensure that $D$ is positive definite to machine precision in practice. These steps include, if necessary, taking the singular value decomposition of the matrix $D = USU'$ where $S$ is a diagonal matrix of singular values. $S$ is clipped to a lower bound of $10^x$, where the resulting matrix is now referred to as $T$. Setting $D = UTU'$ and then $D = (D+D')/2$ ensures symmetry to machine precision. The value of $x$ is chosen to start at -16 and increases by increments of 1 until $D$ is symmetric positive definite to machine precision.

## 4.3. Modifications to the Method

If $Q_1,\ldots,Q_N$ are the amounts that a given entity receives from each of its $N$ direct suppliers, and $Q'_1,\ldots,Q'_N$ are the transaction amounts that the model predicts, it specifically is assumed that $\Delta Q_1,\ldots,\Delta Q_N$ follows a multivariate normal distribution with zero mean and covariance matrix $D$ where $\Delta Q_i = Q'_i - Q_i$ for $i = 1,\ldots,N$.

Following the methodology presented in [21], the authors' method outputs $N + 1$ different combinations of $\Delta Q_1,\ldots,\Delta Q_N$ that would predict the receiving entity achieving an $L$ gain or loss. For each time a reverse stress test is run, all entities were equally weighted to not inadvertently prioritize certain entities over others. In other words, a target entity is willing to receive goods from anyone who can provide it. Other weights can be used, if desired. Mathematically, this means that $w = (100/N)\vec{1}$.

Equations 1-10 in [21] are adopted for the method in this paper, with some modifications driven by the need to determine actual quantities instead of just percent changes. The authors define orthogonal matrix $P$ by "starting with a matrix of full rank with $w$ as a first row." In this paper, matrix $A$ was defined as $A = \text{diag}(w)$ and then the first row of $A$ was set equal to $w$. A $QR$ factorization was performed on $A'$ and $P$ was set to $Q/$. This is mathematically the same as performing the Gram-Schmidt process on the rows of $A$, as the authors instruct to do. It was ensured that the first row of $P$ were all nonnegative values, achieved by multiplying the entire corresponding column by -1 if the first-row entry was negative.

This paper assumes that the weights $w_i$ sum to 1. By multiplying these weights by the authors' result $X_i^j$ in the section "Back to Initial Placement", this makes $w'X^j$ the weighted average of supplier changes from one time period to the next. Alternatively, $X_i^j$ is the percentage change in the aggregate supplier quantity multiplied by each weight $w_i$. This translates to calculating $N+1$ different scenarios $\Delta Q_i^j = w_i X_i^j (S/100)$ for $j = 1,\ldots,N+1$ and for $S = Q_1^j + \ldots + Q_N^j$. Furthermore, $k^j$, as defined in Equation 10 of [21] states that $k^1 = 1$ and $k^j \geq q$ for each $j = 2,\ldots,N+1$.

Taking that $Q_i^{j'} = Q_i^j + \Delta Q_i^j$, an additional measure is taken to ensure that no $Q_i^{j'}$ is ever below zero. This is done by setting any negative $Q_i^{j'}$ to zero and taking the sum of the absolute value of the $Q_i^{j'}$'s that were negative, which is called $E$. For

$$c_i = \frac{Q_i^{j'}}{\sum_{i=1}^{N} Q_i^{j'}},$$

$c_i E$ get added to $Q_i^{j'}$, and thus, $X_i^j = Q_i^{j'}/(w_i S/100)$. If there exists $i$ such that $Q_i^{1'}$ is below zero initially and $Q^{1'}$ is therefore changed by this process, then $Q^{1'}$ will no longer be the "mean" solution and should not have a corresponding $k$ value equal to 1. Setting $d$ to be the value of the conditional density function at $Q^{1'}$ *before* applying these changes and using this as the denominator for calculating the $k$ values for each $Q^j$ *after* applying the changes accounts for this problem. When this scenario occurs, it may break the relationship that $k^j \geq q$. However, it still holds that $k^j \geq 0$. Additionally, the corresponding $N+1$ solutions may not have their corresponding $k$ values in descending order after these changes, so they are reordered so that their $k$ values are in descending order.